\documentclass[preprint,12pt]{elsarticle}

\usepackage{amssymb}
\usepackage{amsmath}

\usepackage{color}
\usepackage{array,multirow}

\journal{Nuclear Physics B}

\begin{document}

\begin{frontmatter}



\title{Parton distribution functions and theory parameters: \\ an NNPDF perspective} 

\author[label1]{Richard D.~Ball}
\author[label2]{Tommaso Giani}
\author[labelJYU,labelHIP]{Felix Hekhorn}
\author[label1]{Jaco ter Hoeve}
\author[label4,label5]{Tanjona R.~Rabemananjara}
\author[label4,label5]{Juan Rojo}
\author[label1]{Roy Stegeman}
\author[label6]{Maria Ubiali}

\affiliation[label1]{organization={The Higgs Centre for Theoretical Physics, University of Edinburgh},
             addressline={JCMB, KB, Mayfield Rd},
             city={Edinburgh},
             postcode={EH9 3FD},
             state={Scotland},
             country={UK}}
             
\affiliation[label2]{organization={Max Planck Institute for Physics},
             addressline={Forschungszentrum Garching, Boltzmannstr. 8},
             city={Garching/Munich},
             postcode={85748},
             country={Germany}}
             
\affiliation[labelJYU]{organization={University of Jyvaskyla, Department of Physics},
             addressline={P.O. Box 35},
             city={University of Jyvaskyla},
             postcode={FI-40014},
             country={Finland}}
\affiliation[labelHIP]{organization={Helsinki Institute of Physics},
             addressline={P.O. Box 64},
             city={University of Helsinki},
             postcode={FI-00014},
             country={Finland}}

\affiliation[label4]{organization={Department of Physics and Astronomy},
             addressline={Vrije Universiteit},
             city={Amsterdam},
             postcode={1081 HV},
             country={The Netherlands}}
             
\affiliation[label5]{organization={Nikhef Theory Group},
             addressline={Science Park 105},
             city={Amsterdam},
             postcode={1098 XG},
             country={The Netherlands}}

\affiliation[label6]{organization={DAMTP, University of Cambridge},
             addressline={Wilberforce Road},
             city={Cambridge},
             postcode={CB3 0WA},
             country={UK}}

\begin{abstract}
Parton Distribution Functions (PDFs) are a key ingredient in theoretical predictions for Large Hadron Collider (LHC) observables 
and play a central role in the extraction of precision Standard Model (SM) and Beyond the SM (BSM) parameters from LHC data.
Recent analyses demonstrate that the determination of fundamental SM parameters such as $\alpha_s(m_Z)$, $m_W$, $m_t$, and $\sin^2\theta_W$ is 
strongly influenced by the choice of input PDFs.
In this contribution, we present the status and challenges of PDF determination from the NNPDF perspective, both in stand-alone fits and 
in joint extractions with (B)SM parameters. We place particular emphasis on results for $\alpha_s(m_Z)$, $m_t$, and Wilson coefficients in the SM Effective Field Theory (SMEFT) framework.
\end{abstract}

\begin{graphicalabstract}
\end{graphicalabstract}

\begin{highlights}
\item Parton Distribution Function determination with the NNPDF methodology    
\item Novel methodologies for the simultaneous determination of the PDFs and (B)SM parameters
\item Simultaneous determination of PDFs and $\alpha_s$
\item Simultaneous determination of PDFs, $\alpha_s$ and top mass
\item Simultaneous determination of PDFs and SMEFT Wilson coefficients
\end{highlights}


\end{frontmatter}



\section{Introduction}
\label{sec:intro}
Theoretical predictions for hadron collider observables depend critically on our understanding of the proton’s internal structure, 
encoded in the Parton Distribution Functions (PDFs)~\cite{Gao:2017yyd,Amoroso:2022eow,Ubiali:2024pyg}. As a result, 
extractions of Standard Model (SM) parameters, such as $\alpha_s(m_Z)$, $m_t$, $m_W$, and $\sin^2 \theta_W$, are unavoidably affected by PDF uncertainties.
Recent LHC analyses illustrate this point: PDF uncertainties are the dominant systematic in extractions 
of $\alpha_s(m_Z)$ from $p_T^Z$ spectra by ATLAS~\cite{ATLAS:2023lhg} and in the CMS $W$ mass measurement~\cite{CMS:2024lrd}. 
Similarly, Higgs coupling measurements are limited by PDF uncertainties in production cross sections~\cite{deFlorian:2016spz,Baglio:2022wzu}.

PDFs also play a central role in Beyond the SM (BSM) searches. High-mass resonance searches probe the poorly 
constrained large-$x$ region~\cite{Beenakker:2015rna}, where uncertainties are significant. This leads to PDF-driven 
differences in key observables like the high-mass Drell-Yan asymmetry~\cite{Ball:2022qtp,Abdolmaleki:2023jvw,Fiaschi:2021sin,Fiaschi:2022wgl,Accomando:2019vqt} 
and the $m_{t\bar{t}}$ tail~\cite{Kassabov:2023hbm}. Moreover, global SMEFT fits (see among many Refs.~\cite{Ellis:2020unq,Giani:2023gfq,Celada:2024mcf,terHoeve:2025gey}), 
which 
often rely on LHC data that are also included in PDF fits, can be biased if BSM effects distort the 
fitted PDFs~\cite{Greljo:2021kvv,Hammou:2023heg,Hammou:2024xuj,Cole:2026eex}.

These considerations highlight the need to systematically account for PDF uncertainties in parameter extractions and, ideally, to perform simultaneous 
fits of both PDFs and (B)SM parameters. Our contribution addresses this challenge using the NNPDF framework.
The structure of this contribution is as follows.
In Sect.~\ref{sec:nnpdf} we review the NNPDF4.0 sets, quantify their performance when compared to Run II LHC observables and 
present the closure tests used to validate the fitting methodologies.
Sect.~\ref{sec:simu-methodologies} describes the methodologies used to extract PDFs together with (B)SM parameters, and presents 
results for $\alpha_s(m_Z)$ and $m_t$, as well as the extension of these analyses to joint fits of the PDFs and SMEFT coefficients.
Finally, in Sect.~\ref{sec:summary} we conclude and briefly explore possible directions for future work.

\section{The NNPDF4.0 PDF sets}
\label{sec:nnpdf}
The NNPDF collaboration has been applying machine learning techniques for the determination of
PDFs and their uncertainties since 2008~\cite{Ball:2008by}, by combining a parametrisation of PDFs 
via a feed–forward neural network with a Monte Carlo importance sampling procedure to propagate the experimental 
and theoretical uncertainties into the space of PDFs.

The latest determinations which belong to the NNPDF4.0 family~\cite{NNPDF:2021njg,NNPDF:2021uiq,NNPDF:2024djq,NNPDF:2024dpb,NNPDF:2024nan} 
include deep-inelastic scattering (DIS), Drell-Yan, Drell-Yan with jet, single-inclusive jet and di-jet, single-top and top-quark pair, 
and prompt-photon measurements. PDFs are parametrised with a single neural network, optimised by
means of gradient descent; hyperparameters, such as those that define the architecture of the neural
network, are determined by means of an automated scan of the space of models~\cite{Carrazza:2021yrg}, see Ref.~\cite{NNPDF:2021njg} for a detailed account of the methodology.

This section starts by reviewing in Sect.~\ref{subsec:theory} the recent theoretical advances. We report in
Sect.~\ref{subsec:pheno} on the comparative study of the performance of NNPDF4.0 with other PDF sets in describing
data not yet included in any PDF determination and in Sect.~\ref{subsec:ct} we discuss the statistical validation of 
PDF uncertainties via closure tests.

\subsection{Theoretical advances}
\label{subsec:theory}

In order to discuss advances in the theoretical framework for the computation of observables entering a PDF fit, 
it is instructive to recall the general setup for how predictions are formulated. For an observable $\sigma$, which
depends linearly on the fitted PDF $f(Q_0)$ at the initial scale $Q_0$, e.g.\ cross sections in DIS, we have
\begin{equation} \label{eq:fact}
    \sigma(Q^2) = \hat{\sigma}(Q^2) E^{(5)}(Q^2 \leftarrow m_b^2) A^{(4)}(m_b^2) E^{(4)}(m_b^2 \leftarrow Q_0^2) f^{(4)}(Q_0^2)
\end{equation}
where $E$ denotes the evolution kernel operator (EKO), i.e.\ the solution of the DGLAP equations, which can be computed from the splitting 
functions $P$; $A$ denotes the massive operator matrix elements (OMEs), which transform the PDF between the 
four- and five-flavour scheme~\cite{Maltoni:2012pa,Bertone:2017djs}, and $\hat{\sigma}$ denotes the partonic matrix element.
The three major ingredients, $P, A$ and $\hat{\sigma}$, are perturbative series in both the strong 
coupling $\alpha_s$ and the electromagnetic coupling $\alpha_{em}$.
The truncation of these series at a given perturbative order leads to a source of theory uncertainty, which we call the 
Missing Higher Order Uncertainty (MHOU).
While a variety of different approaches have been suggested on how to estimate MHOUs~\cite{Lim:2024nsk,Tackmann:2024kci,Kassabov:2022orn,Bonvini:2020xeo,Cacciari:2011ze,Duhr:2021mfd,Ghosh:2022lrf}, 
NNPDF uses traditional scale variations and the theory covariance matrix formalism \cite{Ball:2018lag}  
to incorporate MHOU into PDF extractions~\cite{NNPDF:2019vjt,NNPDF:2019ubu,NNPDF:2024dpb}. 
The shifts in the theory predictions associated with the scale variations can be evaluated using tools that 
produce fast theory predictions via interpolation grids, such as {\tt PineAPPL}~\cite{Carrazza:2020gss,christopher_schwan_2025_15635174}. 

In recent years, significant progress has been achieved in the precision of the perturbative ingredients.
Splitting functions $P$ and the OMEs $A$ are currently known to next-to-next-to-next-to-leading order (N$^3$LO) accuracy in the strong 
coupling $\alpha_s$ and recently PDF extractions at this order have been presented by the NNPDF~\cite{NNPDF:2024nan} and MSHT~\cite{McGowan:2022nag} groups.
A combination of the two PDF determinations has been presented in Ref.~\cite{Cridge:2024icl}, which demonstrates that the use of N$^3$LO PDFs is 
now mandatory for high-precision predictions, such as the total Higgs cross section. 
The combination also includes QED corrections to the splitting functions $P$, which are known at next-to-next-to-leading order (NNLO)
and incorporated into several PDF extractions~\cite{NNPDF:2024djq,Cridge:2021pxm,Xie:2021ajm}. 
An important implication of considering QED corrections is the inclusion of a photonic content of the proton, 
for which the {\tt LuxQED} procedure~\cite{Manohar:2016nzj,Manohar:2017eqh} has become the standard approach.

\begin{figure}
    \centering
    \includegraphics[width=0.44\linewidth]{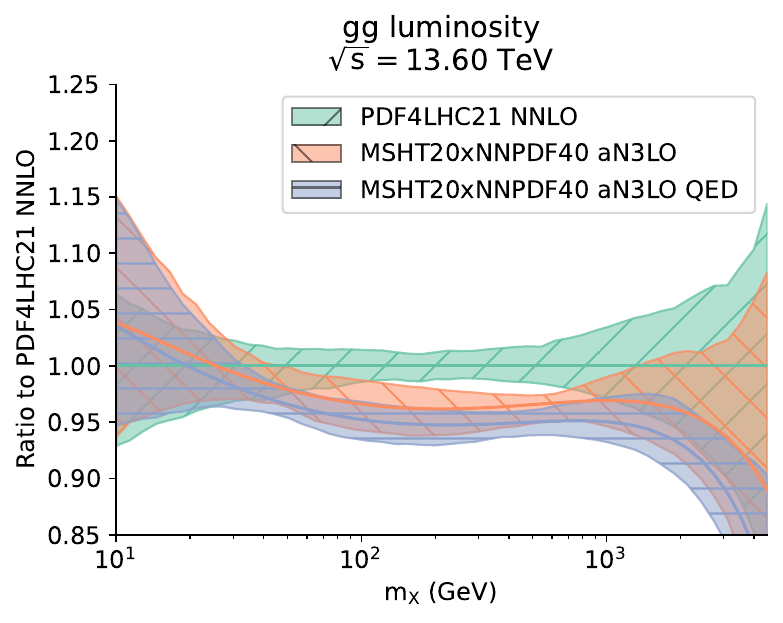}%
    \includegraphics[width=0.48\linewidth]{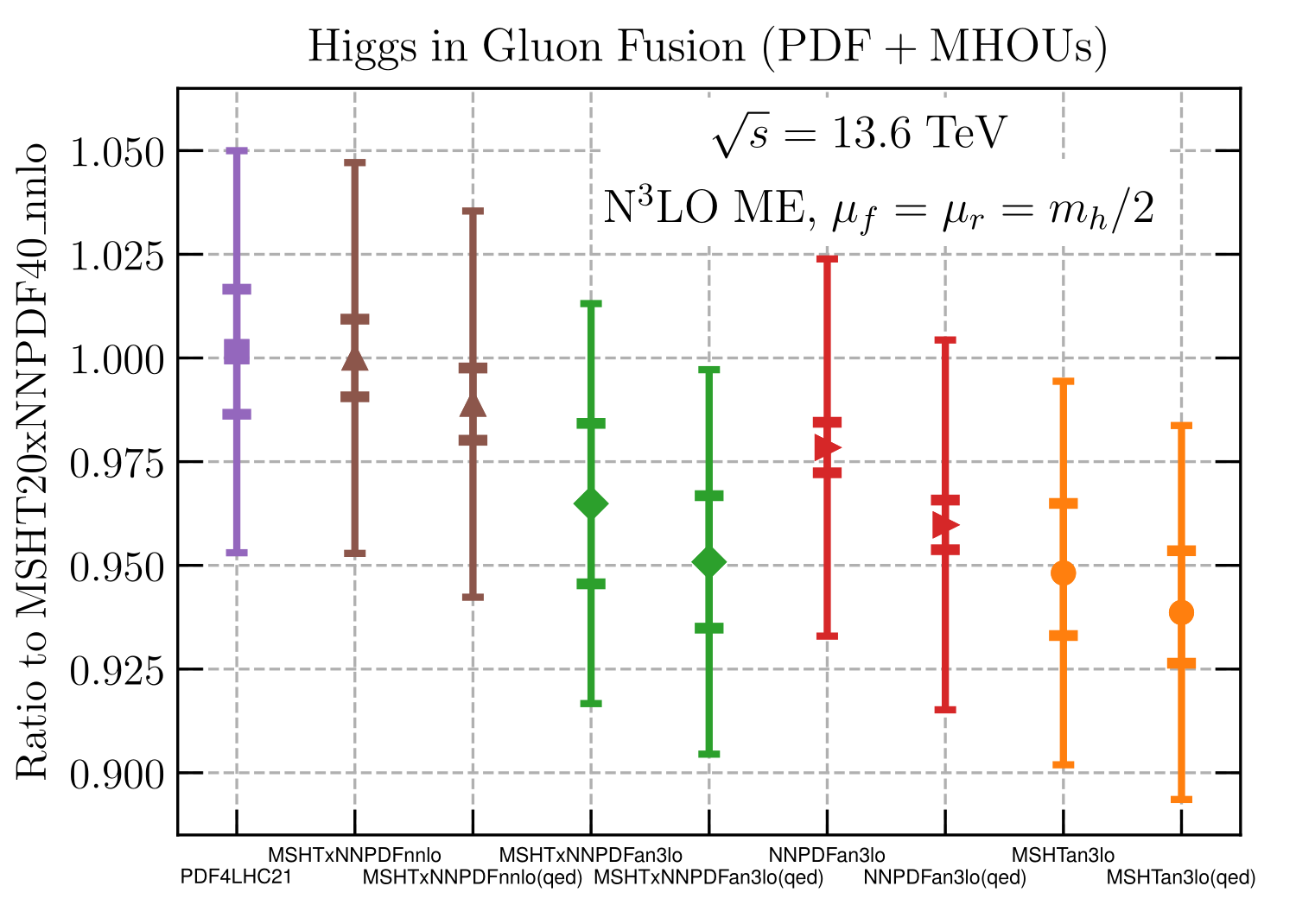}
    \caption{Impact of aN$^3$LO QCD and QED corrections onto the gluon-gluon luminosity at $\sqrt{s}=13.6\,$ TeV (left) and the Higgs cross section in gluon fusion (right) shown as ratio against PDF4LHC21~\cite{PDF4LHCWorkingGroup:2022cjn} (left) or the NNLO combination of Ref.~\cite{Cridge:2024icl} (right).
    On the right panel, the error bars correspond to MHOU combined with PDF uncertainties (outer) or only the latter (inner).}
    \label{fig:theory}
\end{figure}
In Fig.~\ref{fig:theory} we observe that the inclusion of approximate N$^3$LO QCD 
calculations combined with NLO QED corrections 
results in a shift around 2$\sigma$ in the gluon-gluon luminosity and a 
downward shift of the Higgs production cross section in gluon fusion at 13.6 TeV by 3.3\% 
(for the combination of MSHT20 and NNPDF4.0~\cite{Cridge:2024icl}) compared to the NNLO PDFs~\cite{PDF4LHCWorkingGroup:2022cjn} 
used by the Higgs Cross Section Working Group as a baseline. We also note that QED corrections and
the photon PDF~\cite{NNPDF:2024djq,Cridge:2021pxm} have a significant impact.

A second, independent line of improvements is the inclusion of full electro-weak (EW) corrections. 
EW corrections to the partonic matrix elements $\hat\sigma$ are known for several LHC processes~\cite{Armadillo:2024ncf,Armadillo:2024nwk},
but only scarcely for other processes, such as DIS, which in this case even requires a new factorisation approach~\cite{Cammarota:2025jyr}.
A consistent definition of final-state leptons~\cite{Denner:2019vbn}, for example considering bare leptons in both theoretical observables and experimental measurements, 
remains an open challenge for the full inclusion of EW effects in PDF and parameter extractions.


\subsection{NNPDF4.0 performance: a comparative study}
\label{subsec:pheno}

In this section, we summarise the key findings of Ref.~\cite{Chiefa:2025loi}, a systematic comparison of NNLO accurate theory predictions, obtained by using different input
PDF sets (ABMP16~\cite{Alekhin:2017kpj}, CT18, CT18A, CT18Z~\cite{Hou:2019efy},
MSHT20~\cite{Bailey:2020ooq}, NNPDF3.1~\cite{NNPDF:2017mvq},
NNPDF4.0~\cite{NNPDF:2021njg}, PDF4LHC15~\cite{Butterworth:2015oua}, PDF4LHC21~\cite{PDF4LHCWorkingGroup:2022cjn}), for a variety of high-precision experimental measurements that have not yet been
included in any PDF fit. These measurements include differential cross section data from LHC Run II for Drell-Yan gauge boson, top-quark pair, single-inclusive jet and
di-jet production, and from HERA for single-inclusive jet and dijet production.

To assess the agreement between the experimental data and the theoretical predictions the (reduced) $\chi^2$ for each data set 
\begin{equation}
  \chi^{2}
  =
  \frac{1}{n_{\rm dat}}
  \sum_{i, j=1}^{n_{\mathrm{dat}}}
  (T_i-D_i)(C^{-1})_{i j}(T_j-D_j)\,,
  \label{eq:chi2}
\end{equation}
must be computed, where $n_{\mathrm{dat}}$ is the number of data points in the considered dataset,
$D$ are the central values of the experimental data, $T$
are the corresponding theoretical predictions obtained with the central PDF and $C$
is a covariance matrix (defined below). For
agreement between data and theory, one expects $\chi^{2}\sim 1$,
with statistical fluctuations of the order of the standard deviation of the
$\chi^2$ distribution, $\sigma_{\chi^2}=\sqrt{2/n_{\mathrm{dat}}}$. 

In a quantitative assessment of the agreement between theory predictions and experimental data, 
the covariance matrix $C$, in addition to the experimental uncertainties, 
and corrections for nuclear effects in data taken on deuteron or heavy nuclei targets~\cite{Ball:2018twp,Ball:2020xqw}, 
should also include all relevant sources of theoretical uncertainties,
in particular those associated with MHOs, PDFs and
the value of the strong coupling $\alpha_s(m_Z)$. Assuming that all of these
theoretical uncertainties follow a Gaussian distribution and are
mutually independent, they can be incorporated into the covariance matrix
following the formalism developed in~\cite{Ball:2018lag}, in which the covariance matrix in Eq.~\eqref{eq:chi2}
is simply the sum of all the various contributions:
\begin{equation}
\label{eq:total_covmat}
C = C_{\rm exp} + C_{\rm mho} + C_{\rm pdf} + C_{\rm as}.
\end{equation}
The experimental covariance matrix is sometimes provided together with the
experimental measurements; otherwise, in most cases, it can be reconstructed from
knowledge of the correlated and uncorrelated experimental uncertainties.
The contribution to the covariance matrix due to MHOs can be estimated
as the difference between theoretical predictions computed at varied
renormalisation and factorisation scales, $\mu_R$ and $\mu_F$.
In~\cite{Chiefa:2025loi} the 7-point scale variations are treated 
as independent nuisance parameters~\cite{NNPDF:2019vjt,NNPDF:2019ubu,Ball:2021icz}, and the MHO covariance matrix, is built accordingly.  
The contribution to the covariance matrix due to PDF uncertainties is
determined using the definition of covariance between random variables $T_i$ and $T_j$, 
both for Hessian and Monte Carlo PDF sets.
Finally the contribution to the covariance matrix due to the uncertainty of the value
of $\alpha_s(m_Z)$ can be determined consistently with the prescription
of~\cite{PDF4LHCWorkingGroup:2022cjn}, according to which PDF and 
$\alpha_s$ uncertainties are added in quadrature. Taking 
$\alpha_s(m_Z)=0.118\pm 0.001$, consistently with
the latest PDG average~\cite{ParticleDataGroup:2024cfk}, a covariance matrix can be built explicitly from 
the shifts of theoretical predictions computed at $\alpha_s(m_Z)=0.117,0.118,0.119$.
The explicit definitions of all terms appearing in Eq.~\eqref{eq:total_covmat} are given in Ref.~\cite{Chiefa:2025loi}. 

\begin{figure}[tb]
    \centering
    \includegraphics[width=0.8\linewidth]{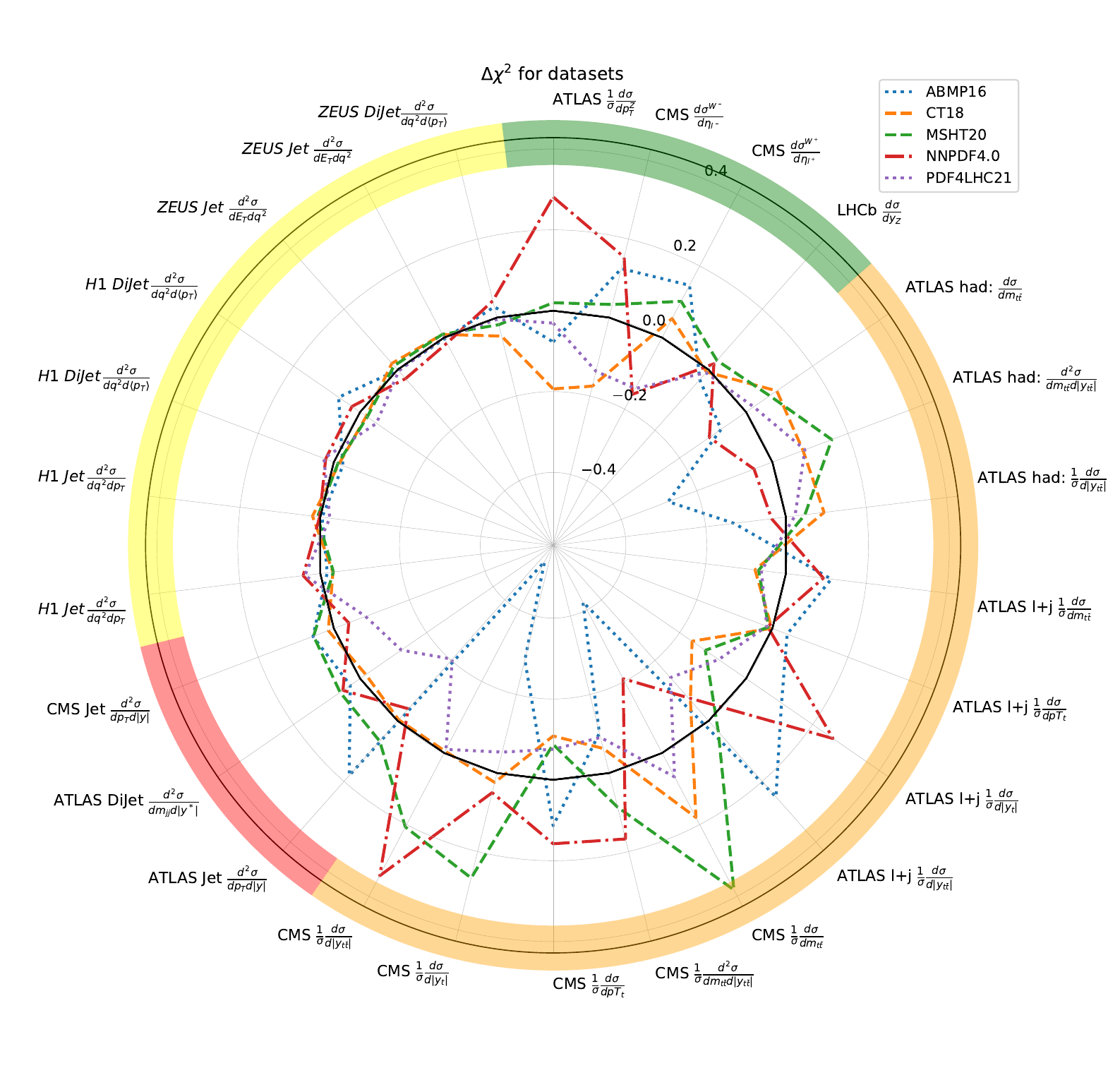}
    \caption{The relative change in the total $\Delta \chi^2$ as given by Eq.~\eqref{eq:delta_chi2}
    for a selection of PDF sets considered in~\cite{Chiefa:2025loi}. The $\Delta \chi^2 = 0$ circumference is represented
    by the solid curve. From Ref.~\cite{Chiefa:2025loi}.}
    \label{fig:delta_chi2}
\end{figure}
The agreement between experimental data and theory predictions, for various 
PDF sets, including all these sources of uncertainty, 
is summarised in Fig.~\ref{fig:delta_chi2}. The estimator used in this plot is  
\begin{equation}
  \label{eq:delta_chi2}
  \Delta \chi^{2(i)}
  =
  (\chi^{2(i)}-\left\langle\chi^{2}\right\rangle_{\rm pdfs})/\left\langle \chi^{2} \right\rangle_{\rm pdfs}\,,
\end{equation}
where the index $i$ runs over the $n_{\rm pdfs}$ input PDF sets considered, and the average is done over all the PDF sets, so that by construction, 
$\sum_i \Delta \chi^{2(i)}=0$. Thus $\Delta \chi^{2(i)}$ is the
relative change in the value of the $\chi^2$ for a given PDF set, compared to the average evaluated over all PDF sets considered in the study. 
%
The conclusion from 
Fig.~\ref{fig:delta_chi2} is 
that, once all sources of uncertainty are taken into account, all of the PDF sets considered in Ref.~\cite{Chiefa:2025loi} 
yield comparable descriptions of all the datasets, despite the fact that some sets have significantly larger PDF uncertainty than others.

\subsection{Tests for robust uncertainty estimate in PDF fits}
\label{subsec:ct}

In the LHC precision era, extracting PDFs and physics parameters requires statistically robust methods and 
reliable uncertainty estimates. Validating PDF uncertainties involves two key aspects: the capacity of 
a fitting methodology to learn the underlying truth, and its generalisation ability. 
While the latter was introduced in~\cite{Cruz-Martinez:2021rgy} and 
tested in the study presented in Sect.~\ref{subsec:pheno}, in this section we focus on the former, using closure tests 
and exploring the impact of experimental inconsistencies.

Following Refs.~\cite{DelDebbio:2021whr,Barontini:2025lnl}, and assuming Gaussian noise, experimental data values $y_0$ can be modeled as
\begin{align}
    y_0 &= f + \eta,
    \label{eq:l1_data}
\end{align}
where $f$ is the true (unknown) observable vector, and $\eta \sim \mathcal{N}(0, C)$ is the noise with covariance matrix $C$, 
which includes both statistical and systematic uncertainties (experimental and theoretical).
A closure test applies the fitting method to synthetic data generated from a known underlying law $\bar{f}$ 
generated from a given PDF set and setting $f=\bar{f}$ in Eq.~\eqref{eq:l1_data}.
By generating $N_{\rm fit}$ such datasets and performing fits, one obtains $N_{\rm fit}$ predicted PDF sets. 
%
To evaluate the quality of uncertainties, Ref.~\cite{Barontini:2025lnl} introduces the normalised bias:
\begin{align}
    \label{eq:bias_definition}
        B^{(l)} & = \frac{1}{n_{\rm dat}}\sum_{i,j=1}^{n_{\rm dat}} (f^{\text{pred}, (l)}_i - \bar{f}_i) (C^{(l)}_{\rm pdf})_{ij}^{-1} (f^{\text{pred}, (l)}_j - \bar{f}_j) ,
\end{align}
which quantifies how far predictions deviate from the truth $\bar{f}$ relative to predicted PDF uncertainties for each fit $l$. 
The root-mean-square normalised bias,
\begin{align}
  \label{eq:bias_variance_ratio_definition}
    R_{b} = \sqrt{\frac{1}{N_{\text{fit}}}\sum_{l=1}^{N_{\text{fit}}} B^{(l)}}\,.
\end{align}
tests whether uncertainties are Gaussian and reliable. Significant deviations of $R_{b}$ from one, corresponding to the expected value for perfectly faithful uncertainties,
are a signal of potential biases introduced by the fitting methodology under consideration.

The study of~\cite{Barontini:2025lnl} also investigates the impact of experimental inconsistencies by generating 
closure test data with the correct covariance matrix $C$, but fitting with a modified matrix where some 
uncertainties are rescaled, simulating underestimated systematics. This mismatch introduces tension between datasets.
The NNPDF methodology is found to remain robust against moderate inconsistencies, producing accurate PDFs and uncertainties that 
generalise well, even to data strongly correlated with inconsistent inputs. Only severe underestimation of systematics 
in systematics-dominated regions significantly distorts results. The authors also propose a data-driven method, based on closure tests, 
to detect such inconsistencies in real-world situation. Such approach, based on a pre-processing of the data that yields a consistent input dataset, 
rather than including an explicit tolerance factors in the fit~\cite{Harland-Lang:2024kvt} is what distinguishes the NNPDF approach from that of 
other global fitting collaborations, such as CT~\cite{Hou:2019efy} or MSHT~\cite{Bailey:2020ooq}. 

Together, these findings demonstrate that as more data are incorporated into global PDF fits, it is possible to maintain both precision and accuracy.

\section{Simultaneous extractions of PDFs and external parameters}
\label{sec:simu-methodologies}

PDFs are extracted by comparing hadronic process data and uncertainties to SM predictions computed in perturbation theory. 
These predictions depend in turn on SM parameters, which are typically held fixed in any particular PDF determination, 
often to values determined independently, through different processes (for example leptonic production, or particle decays).
However, we also want to determine constraints on SM parameters from hadronic data, and the theoretical predictions for these data 
necessarily depend on the PDFs. Indeed, at LHC, 
the main reason we need precise and accurate 
PDFs is that we want to use them to determine SM (and BSM) parameters both accurately and precisely.

To avoid double counting, data can be split into two sets: one set to determine the PDFs (for a particular set of parameters), 
and the other set to determine the parameters (for a particular set of PDFs). 
This works for data which depend sensitively on the parameter in question, but only weakly on the PDFs. 
For example, $gg\to H\to\gamma\gamma$ provides an accurate determination of the Higgs mass, since the position of the diphoton 
peak is insensitive to the background which depends on the PDFs. However for some interesting parameters, such as $\alpha_s$, a 
clean separation is not so easy to achieve. 
For example jet cross sections depend sensitively on both $\alpha_s$ and the gluon, but the $\alpha_s$ dependence is difficult to 
disentangle from the gluon distribution, since they are strongly correlated~\cite{Forte:2020pyp,Forte:2025pvf}.
Another problematic situation is that in which one might want to disentangle new physics effects from the PDFs: 
if all high-energy distributions, where heavy new physics might show, were removed 
from PDF fits and used only for BSM parameter determination, then the large-$x$ gluon, quarks and anti-quarks would be poorly constrained, 
as our knowledge of these distributions would rely on old datasets and on 
the ability of PDF fitters to have good control of the extrapolation region.

Ideally, all data would be used in a global fit to constrain both PDFs and parameters simultaneously, 
accounting for all correlations -- experimental, theoretical, and between PDFs and parameters. 
This is a challenging problem. Aside from the logistic and computational difficulties, 
there is a methodological difficulty. (B)SM parameters form a discrete finite set, so fitting them is in principle very clean. 
PDFs on the other hand are continuous functions, and so their determination is more subtle. 
Thus combining the PDF determination with the parameter determination requires care, to avoid contaminating 
the parameter determination with the dirtier elements of the PDF determination (such as for example tolerance, or cross-validation). 

\subsection{Methodologies for simultaneous PDF and parameter fitting}
\label{subsec:methodologies}

In this section we will describe three distinct methods by which this might be achieved: the correlated replica method 
relying on frequentist Monte Carlo resampling, the Bayesian approach based on theory covariance matrix, and the 
purely numerical optimisation method, {\tt SIMUnet}.


\noindent\paragraph{Correlated Replica Method (CRM)}
In the CRM, data replicas are generated according 
to a multi-gaussian distribution centred on the experimental central values
and with covariance matrix given by the first two terms of Eq.~\eqref{eq:total_covmat}, 
representing the experimental $t_0$-~\cite{Ball:2009qv} and the theory-covariance matrix~\cite{Ball:2018lag}, 
including all experimental, MHOUs~\cite{NNPDF:2019vjt,NNPDF:2019ubu} and other theory uncertainties such as those due to nuclear effects~\cite{Ball:2018twp,Ball:2020xqw}. 
An optimal fit is then determined for each of the $N_{\text{rep}}$ data replica, 
by minimizing a loss function computed on a training subset of data 
and stopping the training conditionally on the loss computed on the remaining
validation subset. By repeating this procedure for a number $N_{\alpha_s}$ of different
values of $\alpha_s$, one finds a vector of best fit results 
for each data replica. This then leads to a Monte Carlo representation 
of the probability distribution in joint $\alpha_s$, PDF space, 
\[
    \{\alpha_s^{(k,h)}\,,\quad k = 1, ...,N_{\text{rep}}\,, \quad h = 1, ...,N_{\alpha_s} \}\,,
\]
from which the most likely value of $\alpha_s$ and associated confidence levels (CL), as well as the 
correlations with the PDFs, can be determined by fitting parabolas to $\alpha_s^{(k,h)}$ for fixed k, and 
minimising to given a replica ensemble $\overline{\alpha_s}^{(k)}$.

\noindent\paragraph{Theory Covariance Method (TCM)}
The TCM can be used to determine analytically the posterior distribution
of any given nuisance parameter $\lambda$ given the data $D$ and the theory predictions $T$.
The nuisance parameter induces a shift on the theory predictions
$T \to T+\lambda \beta$, and assuming Gaussian distributions for both its prior and the likelihood, 
one finds out that the posterior $P(\lambda | T, D)$ is again a Gaussian, 
whose mean $\bar{\lambda}$ and width $Z$ can be written in terms of $D$, $T$ as
\begin{align}
    \label{eq:lambar}
    \bar{\lambda}(T,D)&=\beta^T (C+S)^{-1} (D-T) ,\\
    \label{eq:lamwidth}
    Z &= 1-\beta^T (C+S)^{-1} \beta\,.
    \end{align}
Here $C_{ij}$ is the experimental covariance matrix given in the first term of Eq.~\eqref{eq:total_covmat}, 
and $S_{ij}$ is an additional contribution associated with the nuisance parameter $S_{ij} =  \beta_i\beta_j$.

Performing this determination for each data replica leads to an ensemble 
$\bar{\lambda}^{(k)}$ with expectation value and variance over replicas given by
\begin{align}
    \label{eq:mean_nuisance_parameter}
    &\langle\bar{\lambda}\rangle =
    \beta^T  (C+S)^{-1}(D- T^{(0)}) \, , \\
    &\bar{Z}\equiv 1 - \beta^T (C)^{-1}\beta
+\beta^T (C+S)^{-1} X (C+S)^{-1}\beta  \,,
\end{align}
where $T_i^{(0)} = \langle T_i^{k)}\rangle$ is the theory prediction averaged over replicas, 
while $X_{ij}$ is the covariance matrix $X_{ij} = \langle (T_i^{(k)}-T_i^{(0)})(T_j^{(k)}-T_j^{(0)})\rangle$.
This formalism can for instance be applied to the determination of $\alpha_s$, by defining 
the nuisance parameter as the variation of $\alpha_s$ with respect to 
its prior central value $\alpha_s^{(0)}$
\begin{align}\label{eq:lamalpha}
    \lambda_{\alpha} = \alpha_s - \alpha_s^{(0)}\,,
\end{align}
and by computing $\beta$ by considering a linear expansion of the theory predictions in $\lambda_{\alpha}$.
Generalisation to multiple nuisance parameters, and their correlations, is straightforward \cite{Ball:2021icz}. 

\noindent\paragraph{Simunet Method}
The {\tt SIMUnet} framework \cite{Iranipour:2022iak, Kassabov:2023hbm, Costantini:2024xae}, 
optimises the $\chi^2$ from Eq.~\eqref{eq:total_covmat} by extending the NNPDF4.0 architecture by an extra layer to account for the SMEFT corrections. 
The Wilson coefficients thus enter as trainable parameters living on the edges of the final output layer. 
{\tt SIMUnet}, based on the first tagged version of the public {\tt NNPDF} code~\cite{NNPDF:2021uiq}, augments it with a number of new 
features that allow the exploration of the correlation between a fit of PDFs and BSM degrees of freedom. The public tool allows a global 
analysis of the SMEFT combining the Higgs, top, diboson and electroweak sectors, and more data can be added in a rather straightforward 
way to combine the sectors presented here alongside Drell-Yan and flavour observables, for example. {\tt SimuNET}, although developed specifically for SMEFT, 
can be extended to the simultaneous fit of SM parameters. 

\subsection{Testing the methodologies for simultaneous PDF and parameter fitting}
\label{subsec:ctsimu}
The formalism of closure tests discussed in Sect.~\ref{subsec:ct} can also be used to validate a simultaneous extraction of external parameters and PDFs. 
Synthetic data are generated according to a given underlying law, represented by
a known set of PDFs together with a chosen value of the parameters. 
The methodology is then run on these data,
and the procedure is repeated $N_{\text{fit}}$ times, i.e. $N_{\text{fit}}$ sets of data are generated, 
leading to the determination of $N_{\text{fit}}$ different values of the parameters.
Just as for for PDFs, statistical estimators can be defined to assess the reliability of the results. 

For example, in a determination of $\alpha_s$, the analogue of Eq.~\eqref{eq:bias_definition} is given by
\begin{align}
   \label{eq:alphas_Rbvk}
    \mathcal{R}_{\rm  bv}^{(l)}=  \frac{\alpha_s^{(l)} - \bar{\alpha}_s}{\sigma_\alpha^{(l)}},
\end{align}
where $\bar\alpha_s$ is the true underlying value of $\alpha_s$, and
$\alpha_s^{(l)}$ and $\sigma_\alpha^{(l)}$ are, respectively, the
central value and uncertainty of $\alpha_s$ obtained in the $l$-th fit. 
The normalised bias can be computed exactly as in Eq.~\eqref{eq:bias_variance_ratio_definition}, 
and the fitting methodology tested against possible biases by comparing the resulting value of $\mathcal{R}_{\rm bv}$ with the expected value of one. 
\subsection{Simultaneous determination of PDFs and $\alpha_s$}
\label{subsec:pdf_as}

A key application of the simultaneous fitting methodologies that we have listed so far is the simultaneous determination of 
PDFs and $\alpha_s(m_Z)$~\cite{Ball:2018iqk,Forte:2020pyp}. The combined PDF+$\alpha_s$ is the dominant source of uncertainty in many PDF analyses, 
and to achieve the goal of one percent accuracy of the theoretical predictions across a large range of 
collider observables, a more precise knowledge of PDFs and $\alpha_s$ is required. In a recent paper~\cite{Ball:2025xgq} based on NNPDF4.0~\cite{NNPDF:2021njg}, 
improved accuracy and precision is achieved thanks to a combination of global dataset, as well as improved methodology and theory inputs.
The determination benefits from the NNPDF4.0 dataset, the largest global dataset entering the 
simultaneous PDF+$\alpha_s$ extraction to date.
In a previous determination based on NNPDF3.1~\cite{Ball:2018iqk} the dominant source of uncertainty was the MHOU, which was estimated as half of 
the shift in $\alpha_s$ obtained at NLO and NNLO. Accounting for MHOUs via the inclusion of a theory covariance matrix, 
along with the inclusion of NLO QED corrections, a photon PDF and approximate N$^3$LO (aN$^3$LO) QCD corrections makes the new determination 
superior in term of theoretical accuracy. 

\begin{figure}[tb]
    \centering
    \includegraphics[width=0.49\linewidth]{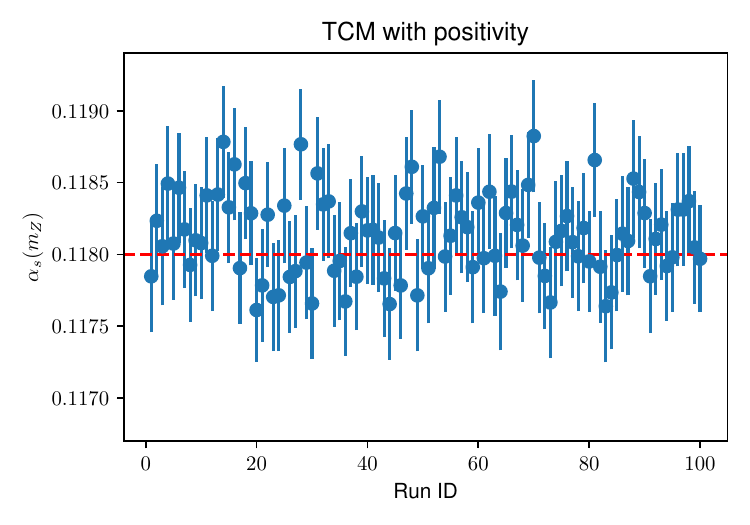}
    \includegraphics[width=0.49\linewidth]{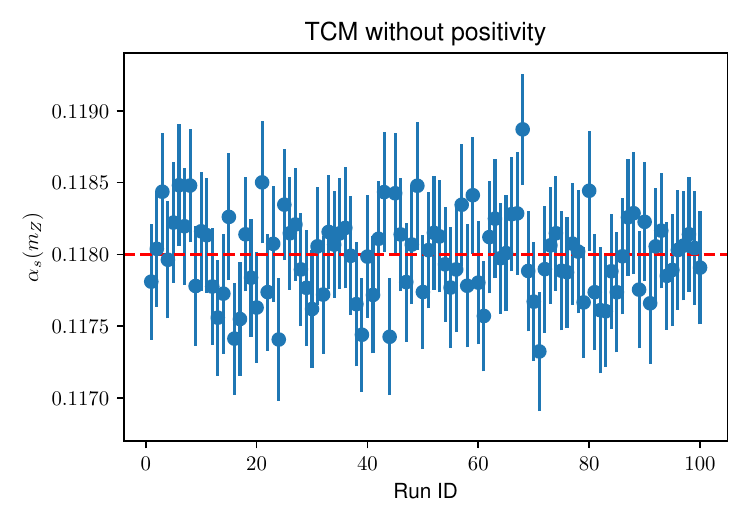}
    \caption{The results of two multi-closure tests consisting of 100 pseudodata samples, performed using the TCM~\cite{Ball:2021icz}, 
    to pseudodata generated at $\alpha_s(m_Z)=0.118$ (denoted by the dashed red line). Left: with positivity, right: without positivity (see text).}
    \label{fig:closure_test}
\end{figure}
In the NNPDF framework, theory predictions are stored in precomputed {\tt PineAPPL} grids at fixed values of $\alpha_s(m_Z)$,  
because re-computing the DGLAP running is not computationally feasible at every training step. 
As a consequence, it is not convenient to treat $\alpha_s$ simply as another trainable model parameter, thus making the application of the {\tt SIMUnet} methodology not so straightforward. 
Nevertheless, the CRM method and TCM method described in the previous section can be both applied in a rather straighforward way to the problem at hand.  
They can be both rigorously tested through a multi-closure test (see Sect.~\ref{subsec:ctsimu}). 
Synthetic data can be generated with $\overline{\alpha}_s(m_Z)=0.118$, and $\alpha_s(m_Z)$ 
extracted from 100 random samples of synthetic data. The results of the simultaneous fit are then compared with the underlying law, 
thereby providing a stringent test of the accuracy of the methodology used. 
For example in the left plot of Fig.~\ref{fig:closure_test} we show results of 100 $\alpha_s(m_Z)$ determinations obtained using the TCM and the standard settings of NNPDF4.0. 
The weighted mean of these determinations is $\alpha_s(m_Z)=0.11813(4)$, and fails the closure test as indicated by the significant disagreement with the input value $\overline{\alpha}_s$. 
The same result is found using the CRM.
The failure of this test is explained by the requirement of PDF and observable positivity in NNPDF4.0, which necessarily breaks the assumed Gaussianity of both experimental and theoretical uncertainties. Releasing these constraints, we find the results shown in the right plot of Fig.~\ref{fig:closure_test} where the weighted mean, $\alpha_s(m_Z)=0.11798(4)$, is in agreement with $\overline{\alpha}_s$. This residual bias is taken into account in the final result.
This illustrates the importance of closure testing, since without these tests, the bias present in the methodology would have been missed. 

\begin{figure}[tb]
    \centering
    \includegraphics[width=0.49\linewidth]{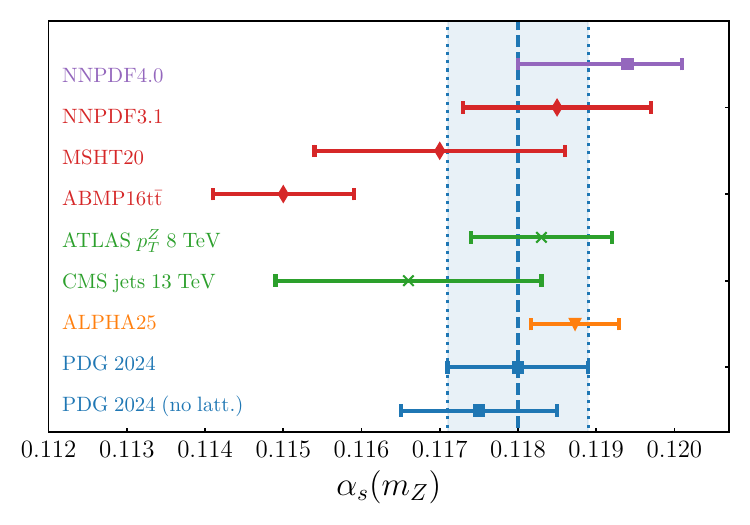}
    \includegraphics[width=0.49\linewidth]{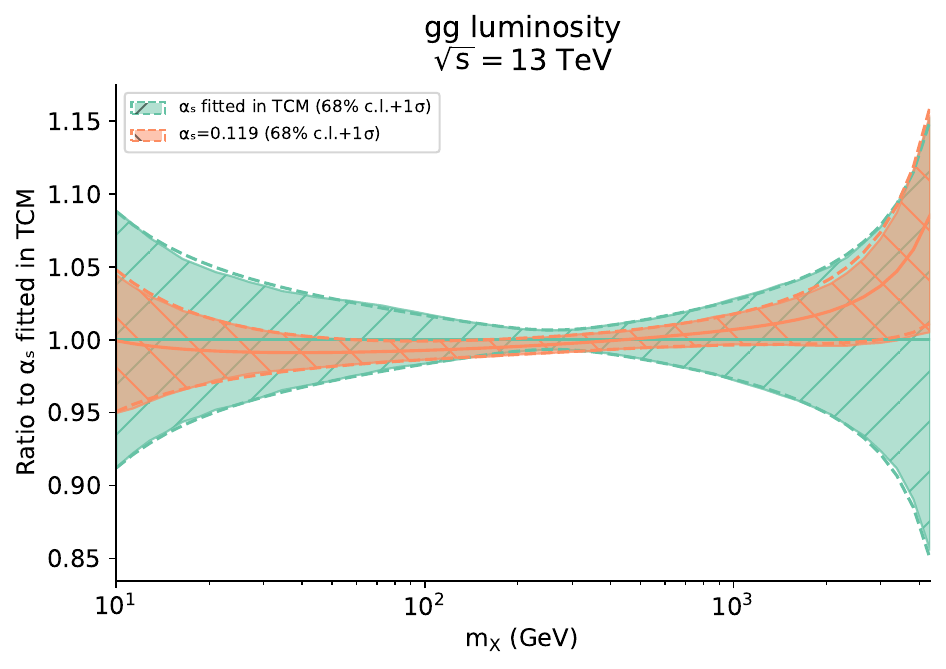}
    \caption{Left panel: A comparison of $\alpha_s(m_Z)$ determinations by, from top to bottom, NNPDF4.0~\cite{Ball:2025xgq}, NNPDF3.1~\cite{Ball:2018iqk}, MSHT20~\cite{Cridge:2024exf}, ABMP16tt~\cite{Alekhin:2024bhs}, ATLAS~\cite{ATLAS:2023lhg}, CMS~\cite{CMS:2021yzl}, ALPHA25~\cite{Brida:2025gii}, and the PDG24~\cite{ParticleDataGroup:2024cfk}
    Right panel: A comparison of gluon-gluon luminosities at $\sqrt{s} = 13,\text{TeV}$ from fits using fixed $\alpha_s(m_Z) = 0.119$ and those including a theory covariance matrix, 
    constructed by varying $\alpha_s(m_Z)$ up and down by 0.002 around 0.119.
    }
    \label{fig:nnpdf40alphas_summary}
\end{figure}
Combining all the aforementioned improvements, a value of $\alpha_s(m_Z) = 0.1194^{+0.0007}_{-0.0014}$ is extracted, 
where the downward uncertainty includes both the experimental and theoretical uncertainties, and the positivity bias estimated by comparing results with and without positivity.
In Fig~\ref{fig:nnpdf40alphas_summary} (left panel) this result is compared to various other determinations. The results are mostly consistent with the PDG value. 
On the right panel the gluon-gluon luminosity from a fit at a fixed value of $\alpha_s$ is compared to a fit in which an $\alpha_s$ covariance matrix is included, 
thus showing that the inclusion of $\alpha_s$ uncertainties in a correlated way with respect to PDF uncertainties leads to a more faithful and robust gluon PDF uncertainty. 
%

\subsection{Simultaneous determination of PDFs, $\alpha_s$ and $m_t$}
\label{subsec:pdf_as_mt}

The TCM methodology applied in Sect.~\ref{subsec:pdf_as} to determine $\alpha_s(m_Z)$ can be applied similarly to a joint determination of $\alpha_s(m_Z)$ and $m_t$. 
An increase in the value of $\alpha_s$ can naively be compensated at the level of the cross section by higher values of $m_t$, suggesting that any individual 
extraction could underestimate uncertainties. So far, simultaneous extractions of $\alpha_s$, $m_t$ and the PDFs have been performed in \cite{Alekhin:2017kpj} 
at NNLO QCD (excluding $t\bar{t}$ differential data), in \cite{Garzelli:2020fmd} at NLO QCD (including $t\bar{t}$ differential data), and, more recently, 
in \cite{Alekhin:2024bhs} while taking into account, for the first time, double differential $t\bar{t}$ at NNLO QCD. Examples of determinations of $\alpha_s$ 
and $m_t$ with fixed PDFs can be found  in \cite{Cooper-Sarkar:2020twv, Cridge:2023ztj}.

\begin{figure}[bt]
    \centering
    \includegraphics[width=0.5\linewidth]{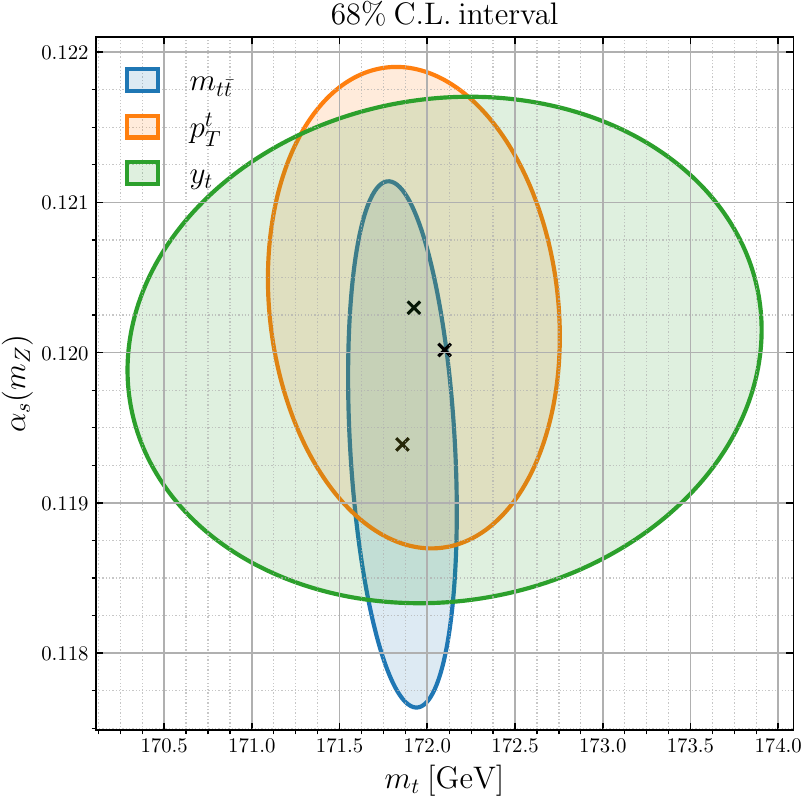}
    \caption{68 \% C.L. intervals in the joint ($\alpha_s, m_t$) plane comparing the impact of including distributions differential in either the invariant mass $m_{t\bar{t}}$ of the top-pair, the top quark transverse momentum $p_T^t$ or its rapidity $y_t$ at NNLO QCD + MHOU accuracy. Note that these results are preliminary.}
    \label{fig:as-mt-tcm-fit}
\end{figure}
Here we focus instead on how $m_t$ and $\alpha_s$ can be determined simultaneously with the PDFs using the TCM formalism outlined in Sect.~\ref{subsec:methodologies}. 
This proceeds by introducing an additional nuisance parameter $\lambda_{m_t}$, defined as the variation of $m_t$ with respect to its prior central value $m_t^{(0)}$, $\lambda_{m_t} = m_t - m_t^{(0)}$, 
so that theory predictions are shifted by an amount $\lambda \beta$ with $\beta$ computed by considering a linear expansion of the theory predictions in $\lambda_{m_t}$.
Fig.~\ref{fig:as-mt-tcm-fit} shows the $68\%$ C.L. exclusion contours in the $(\alpha_s, m_t)$ plane originating from a PDF fit at NNLO QCD + MHOUs 
in such a formalism. We compare the impact coming of including the distributions in the invariant mass $m_{t\bar{t}}$, or the distributions 
in top-quark transverse momentum $p_T^t$, or the  rapidity $y_t$ distributions, noting that these distributions are not independent.  
Double differential distributions in $m_{t\bar{t}}$ and $y_t$ are included in both the $m_{t\bar{t}}$ and $y_t$ determinations. 
We find consistent results with all three distributions. The top mass is determined most precisely from the $m_{t\bar{t}}$ distribution, 
 followed by the $p_T^t$ and $y_t$ distributions, all with limited correlation to $\alpha_s(m_Z)$.

Finally, we remark that the nuisance parameter formalism can be trivially extended to a simultaneous global determination of the other 
heavy quark masses $m_c$ and $m_b$, to a global determination of the mass $m_W$ of the $W$-boson and $\sin^2\theta_W$, or indeed to constrain BSM effects, 
see Sect.~\ref{subsec:pdf_smeft}.

\subsection{Simultaneous determination of PDFs and SMEFT Wilson coefficients}
\label{subsec:pdf_smeft}
In the presence of BSM effects, the partonic matrix element $\hat\sigma(Q^2, c)$ from Eq.~\eqref{eq:fact} may carry an additional dependence on BSM 
parameters $c$. So far we have discussed the extraction of SM parameters and PDFs assuming $c = 0$. However, 
as high-energy data from the LHC are increasingly incorporated into PDF  fits, the tails of the distributions relevant for PDF determination may 
be influenced by potential effects from BSM physics, which may in fact distort the PDFs, 
the strong coupling constant or the top mass. This motivates the realisation of a more general framework where SM and BSM parameters are fitted simultaneously. 
In the remainder of this section we focus on the SMEFT framework at dimension-six to parametrise BSM effects~\cite{Brivio:2017vri}. 

In the SMEFT, the SM partonic matrix element receives higher power corrections in the Wilson coefficients $c$ suppressed by a new physics scale $\Lambda \gg v$ well above the electroweak scale $v$,
\begin{equation}
 \hat\sigma(Q^2, c) = \hat\sigma(Q^2, 0) + \frac{1}{\Lambda^2}\sum_i \hat\sigma^{\rm (eft)}_i(Q^2)c_i  + \frac{1}{\Lambda^4}\sum_{i,j} \hat\sigma^{\rm (eft)}_{ij}(Q^2)c_i c_j \, ,
\end{equation}
where $\hat\sigma^{\rm (eft)}_i$ originates from the interference between the SM and the EFT, while $\hat\sigma^{\rm (eft)}_{ij}$ corresponds to the EFT correction squared. A joint optimisation of the PDF parameters $\theta$ and the Wilson coefficients can then be realised by minimising the figure of merit, Eq.~\eqref{eq:chi2}, 
where the theory predictions $T$ are obtained afrom convolving the partonic matrix element with the PDFs, and depend both on the PDF and on the Wilson coefficients.

Several studies have addressed this issue from both experimental and theoretical 
perspectives~\cite{Kassabov:2023hbm,Greljo:2021kvv,Hammou:2023heg,Cole:2026eex,Costantini:2024xae,CMS:2021yzl, ZEUS:2019cou,Carrazza:2019sec,
McCullough:2022hzr,Gao:2022srd,Shen:2024uop}. Here, we only discuss two examples that are relevant in the context of PDF fitting.
The top quark sector was studied in Refs~\cite{Kassabov:2023hbm,Cole:2026eex}, and it was shown that the bounds of the dim-6 SMEFT operators constraining the top sector 
remain relatively stable once the SMEFT Wilson coefficients are fitted simultaneously with the PDFs. 
However, the gluon PDF itself becomes noticeably softer at large-$x$ when top quark data are included and the Wilson coefficients are fitted simultaneously with the PDFs. 
This underscores a non-trivial interplay between large-$x$ gluon distributions and potential BSM signals in the tails of $t\bar{t}$ invariant mass distributions 
and possibly in high-$E_T$ jet spectra. Although the constraints on the Wilson coefficients are largely robust, the corresponding partonic 
luminosities — especially the gluon-gluon channel — may still exhibit significant shifts.

\begin{figure}[htb]
    \centering
     \includegraphics[width=0.7\textwidth]{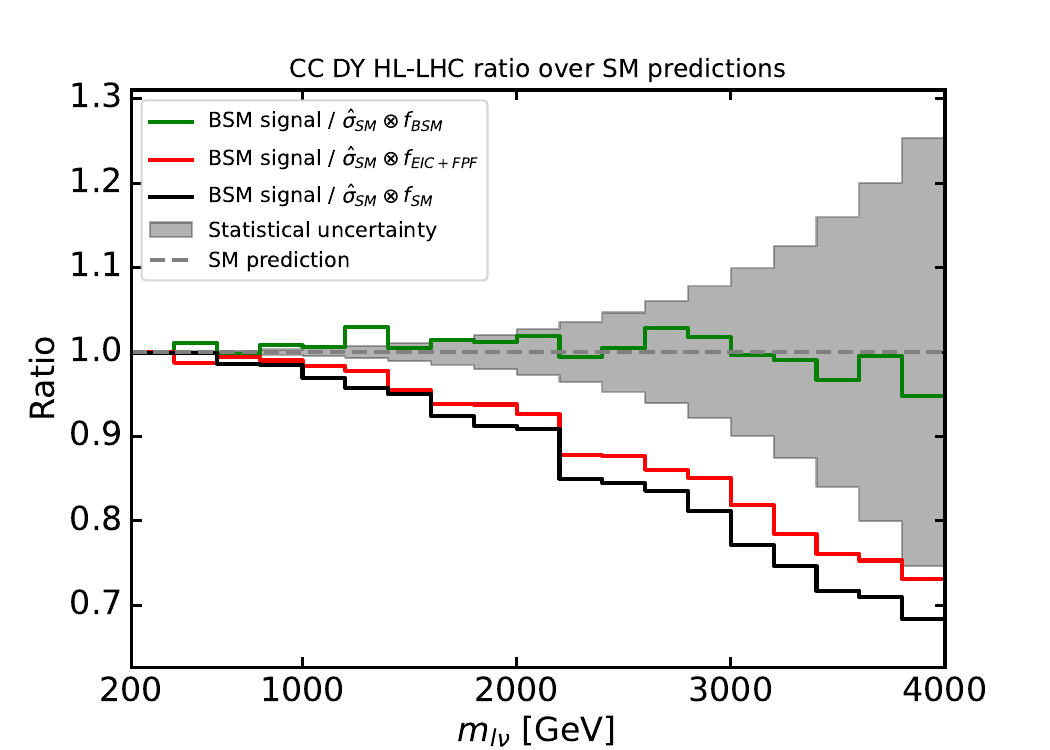}
    \caption{Predictions for charged-current DY inclusive cross
sections differential in $M_l\nu$, with $l = (e, \mu)$ in the presence of a BSM signal associated to 
a flavour-universal $W'$ of $M_{W'}=13.8$ TeV. The distribution is
compared to the SM predictions obtained with a
BSM-biased set of PDFs including only HL-LHC data (green line) or with a set of PDFs
obtained from a joint fit of the HL-LHC data and the
FPF+EIC data (red line), or finally with the "true" SM PDFs
(black line).  From Ref.~\cite{Hammou:2024xuj}} \label{fig:smeftpdf}
\end{figure}
In the context of high-mass Drell–Yan production, on the other hand, 
particularly under the HL-LHC scenario, neglecting the interplay between large-$x$ PDFs and SMEFT contributions in the distribution 
tails could result in missed or misinterpreted signs of new physics. For instance, the inclusion of universal operators 
such as $\hat{W}$ and $\hat{Y}$~\cite{Torre:2020aiz}, 
parametrising flavour-universal heavy $W'$ and $Z'$ respectively, can lead to significantly looser constraints on SMEFT operators if these effects are absorbed into the PDFs~\cite{Greljo:2021kvv,Iranipour:2022iak}. 
Several studies~\cite{Hammou:2023heg,Hammou:2024xuj,Cole:2026eex} demonstrate that universal BSM effects in Drell–Yan tails can be accommodated by softer 
large-$x$ antiquark distributions—far exceeding standard PDF uncertainties. 
This highlights that a viable strategy to mitigate such contamination involves supplementing PDF fits with additional 
low-energy observables, such as those provided by the Electron Ion Collider~\cite{AbdulKhalek:2021gbh} or the Forward Physics Facility~\cite{Feng:2022inv,Cruz-Martinez:2023sdv}, underlining the complementarity of low- and high-energy experimental data. This is highlighted in Fig.~\ref{fig:smeftpdf}, in which the data, in which a BSM signal associated with a new heavy $W'$ boson with
$M_{W'} = 13.8$ TeV and $g'=1$ is injected, is compared to the SM theoretical predictions obtained with a BSM-biased set of PDFs (green
line) or with a set of PDFs obtained from a joint fit of the HL-LHC data and the FPF+EIC data, in which the
HL-LHC data that feature an inconsistent behaviour with the bulk of the data get flagged and excluded from the fit
(red line), or finally with the true SM PDFs (black line). One can see that the green line closely matching the SM prediction
demonstrates that the BSM-biased PDFs mimic the BSM signals present in the data and erase the deviation from
the SM that should be observed. As a result, using these PDFs in theoretical predictions would hide the BSM signal. 

The methodologies discussed in Sect.~\ref{sec:simu-methodologies} provide powerful tools for investigating the relationship between PDFs and new physics effects. 
For instance, {\tt SIMUnet} enables simultaneous fits of SMEFT operators and PDFs, or the injection of any new physics model into 
the dataset, to assess whether its effects may be absorbed in a global PDF fit. The newly released {\tt Colibri} tool~\cite{Costantini:2025agd} allows to generalise 
the idea of {\tt SIMUnet} to a variety of PDF models (from polynomial parametrisations, to Gaussian processes~\cite{Candido:2024hjt} 
and linear ensembles of randomly inizialised neural networks~\cite{Costantini:2025wxp}) and compute joint Bayesian posteriors in the PDF and parameter space.
Alternatively, the TCM method offers a straightforward way to incorporate BSM contributions by introducing a nuisance 
parameter for each Wilson coefficient. As illustrated in Sect.~\ref{subsec:pdf_as}–\ref{subsec:pdf_as_mt}, this framework naturally treats 
SM and BSM parameters — such as $m_t$, $\alpha_s$, and the Wilson coefficients — on an equal footing, without incurring additional computational complexity.

\section{Summary and outlook}
\label{sec:summary}

As the quantity and quality of available LHC measurements continue to improve, ramping up to the HL-LHC high-statistics era, the importance of a careful understanding of PDF systematics for the extraction of (B)SM parameters cannot be underestimated. 
In this contribution, we have reviewed the methodologies used in the NNPDF framework to carry out joint determinations of the PDFs and (B)SM parameters, and highlighted recent results for joint fits of the PDFs together with $\alpha_s(m_Z)$ and $m_t$ (SM) and together with Wilson coefficients of the SMEFT (BSM).
Emphasis is put on faithful uncertainty estimate and statistical validation by means of closure tests. 

The strategies presented in this contribution could be extended in several possible directions.
First, in the same manner as PDFs are fitted together with $\alpha_s(m_Z)$ and/or $m_t$, one could study a joint extraction of PDFs with $m_c$ and $m_b$ along the lines of~\cite{xFitterDevelopersTeam:2016myj}. Second, PDFs play a crucial role in the determination of $m_W$ and $\sin^2\theta_W$ at the LHC, and hence a relevant exercise would be to simultaneously determine PDFs incorporating the same observables that provide sensitivity to $m_W$.
Third, simultaneous determinations of PDFs and SMEFT coefficients are particularly interesting in the context of future collider studies such as the EIC, FPF, LHeC or FCC~\cite{AbdulKhalek:2021gbh,Cruz-Martinez:2023sdv,Bissolotti:2023vdw,LHeC:2020van,Mangano:2022ukr}.
Finally, one may want to extend the NNPDF methodology to unbinned or detector-level observables, which exhibit optimal sensitivity both to the determination of the PDFs~\cite{Aggarwal:2022cki} and to the extraction of (B)SM parameters~\cite{GomezAmbrosio:2022mpm,Benato:2025rgo}.



\bibliographystyle{elsarticle-num-names} 
\bibliography{nnpdf-report.bib}

\end{document}